\documentclass{article}
\usepackage{spconf,amsmath,graphicx}
\usepackage[hidelinks]{hyperref}
\usepackage{amsmath,amssymb,amsfonts}
\usepackage{cite}
\usepackage{svg}
\usepackage[caption=false,font=footnotesize]{subfig}
\usepackage{dblfloatfix}

\title{Towards Real-Time Human--AI Musical Co-Performance: Accompaniment Generation with Latent Diffusion Models and MAX/MSP}
\name{
\parbox{\linewidth}{\centering
Tornike Karchkhadze, Shlomo Dubnov 
}}
\address{
         University of California San Diego \\
}

\begin{document}
\ninept
\maketitle

\begin{abstract}

We present a framework for real-time human--AI musical co-performance, in which a latent diffusion model generates instrumental accompaniment in response to a live stream of context audio. The system combines a MAX/MSP front-end—handling real-time audio input, buffering, and playback—with a Python inference server running the generative model, communicating via OSC/UDP messages. 
This allows musicians to perform in MAX/MSP — a well-established, real-time capable environment — while interacting with a large-scale Python-based generative model, overcoming the fundamental disconnect between real-time music tools and state-of-the-art AI models. 
We formulate accompaniment generation as a sliding-window look-ahead protocol, training the model to predict future audio from partial context, where system latency is a critical constraint. To reduce latency, we apply consistency distillation to our diffusion model, achieving a $5.4\times$ reduction in sampling time, with both models achieving real-time operation.
Evaluated on musical coherence, beat alignment, and audio quality, both models achieve strong performance in the Retrospective regime and degrade gracefully as look-ahead increases. These results demonstrate the feasibility of diffusion-based real-time accompaniment and expose the fundamental trade-off between model latency, look-ahead depth, and generation quality that any such system must navigate.

\end{abstract}

\begin{keywords}
Real-time Accompaniment Generation, Human--AI Co-Performance, Latent Diffusion Models, MAX/MSP
\end{keywords}

\section{Introduction}
\label{sec:intro}

Music is inherently a performative art-form. For most of human history—long before the relatively recent invention of recording technologies—music, an act of realization in sound, existed only in live, performative, and ephemeral contexts~\cite{small1998musicking, cook2021music}. Performative musicianship, whether in the form of improvisation, jamming, or following a known pattern, demands not only technical mastery over an instrument but also coordination, coherence, and interaction with other performers. This coherence spans rhythmic, harmonic, and structural dimensions, and crucially, it also involves anticipation—a continuous, predictive awareness of what is about to happen next. In ensemble performance, musicians mutually anticipate each other's phrases while trusting that all performers will stay within a shared harmonic and rhythmic framework. Musical performance is thus a collaborative, socially embedded practice, involving trust, timing, and mutual responsiveness among human agents~\cite{Keller2008, wrigley2013experience}. With recent advances in artificial intelligence and machine learning systems increasingly taking on creative and generative roles in music, the fundamental question of what form musical live performance takes when some—or all—of the agents involved are machines becomes pressing — yet existing approaches remain far from this goal.

\begin{figure}[t]
  \centering
  \includegraphics[width=1.0\linewidth]{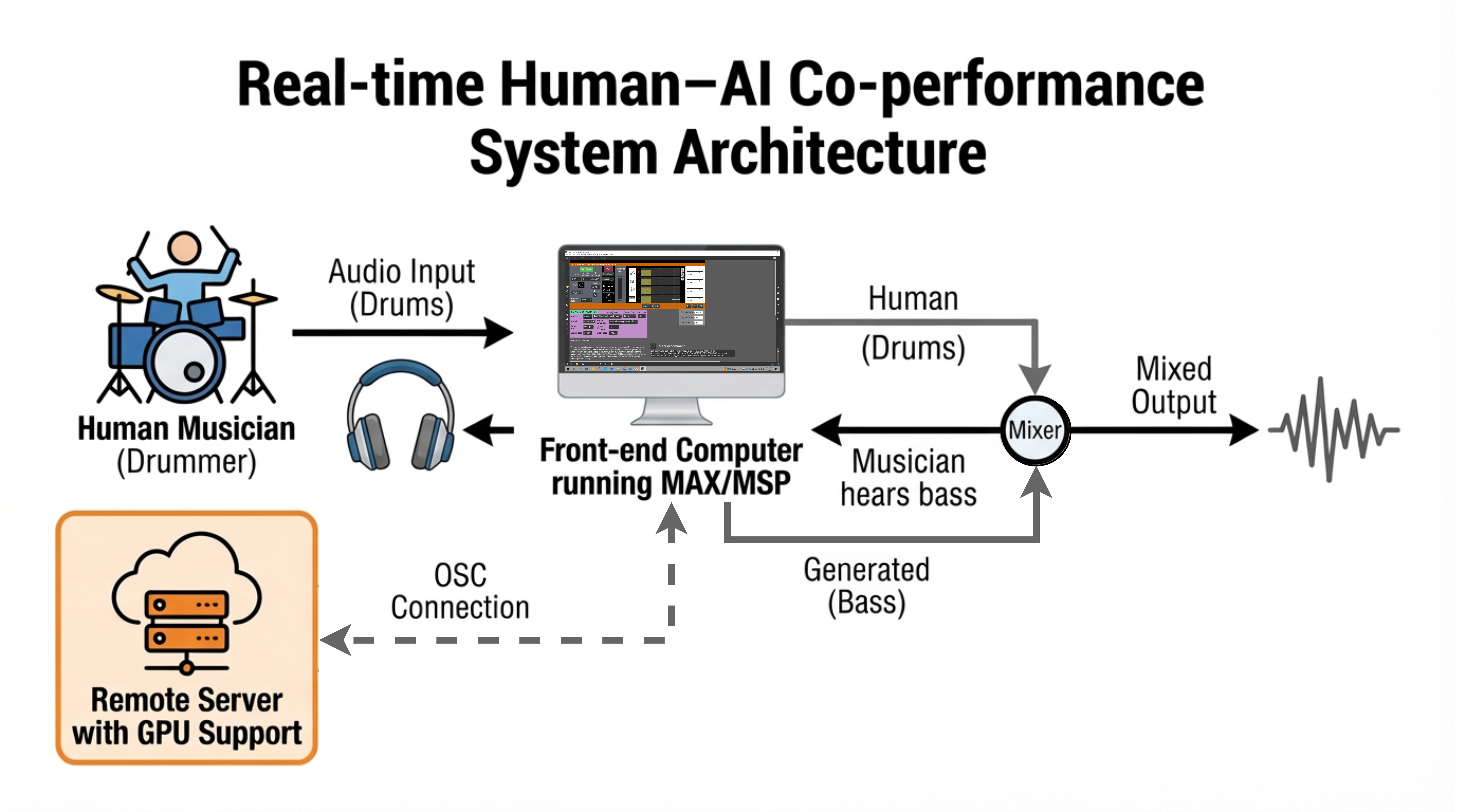}
  \caption{\textbf{Real-time human--AI co-performance system architecture.}
    A musician performs live while MAX/MSP captures the audio stream and communicates with a remote GPU server via OSC/UDP. The server runs a diffusion-based generative model that produces a complementary accompaniment (e.g., bass for a drummer), which is returned and mixed with the live performance in real time.}
  \label{fig:system_overview}
\end{figure}

Most modern machine learning models for music generation have primarily focused on text-to-music synthesis, where a prompt is interpreted as a complete musical piece~\cite{agostinelli2023musiclm, dhariwal2020jukebox, MusicGen2023, chen2024musicldm}, and are not suitable for co-performative formats. Other systems that aim to generate accompaniment~\cite{parker2024stemgen, donahue2023singsong, rouard2025musicgen, karchkhadze2024multitrack, karchkhadze2025Simultaneous} typically operate in offline/retrospective mode, where the full input must be provided before the accompaniment can be generated. 
These systems are typically large-scale generative models with considerable inference latency. In standard digital audio, sound is processed in small, continuously replenished buffers — typically a few milliseconds long — and audio must be ready before each buffer begins. This works because processing is fast enough to keep up. Generative model inference, however, is orders of magnitude slower than a single such buffer, so audio cannot be generated on demand.
To overcome this, similar to how human musicians anticipate, a model must generate ahead of playback, building up a enough buffer that can be consumed while the next prediction is computed, thus compensating for its own inference delay.
The need for look-ahead makes accompaniment generation difficult for existing models, which are not designed for such conditions and typically yield misaligned, incoherent output when forced into a Look-ahead regime.

A further obstacle for real-time human--AI musical performance lies in the disconnect between machine learning infrastructure and musician-facing environments. State-of-the-art generative models are implemented in Python and typically run on remote GPU servers not designed for low-latency musical interaction. Additionally, Python does not natively support real-time audio processing or synchronization with external audio hardware. As a result, even if capable models were available, there is no standardized way for musicians to interface with them directly—plugging in an instrument and receiving a musically aligned response in real time. Taken together — the need for look-ahead generation and the lack of a musician-facing interface — real-time musical accompaniment with AI models remains largely an open problem.

In this work, we introduce a framework for real-time instrumental accompaniment based on large generative models. We develop a Latent Diffusion Model (LDM)~\cite{liu2023audioldm, chen2024musicldm} trained under denoising score matching~\cite{song2019generativemodeling} paradigm. The model uses the Music2Latent~\cite{Pasini2024Music2Latent} encoder-decoder to operate in a compressed latent space, where it is conditioned on a mixture of input tracks to generate a requested instrumental stem. We train on four instruments—bass, drums, guitar, and piano—using the Slakh2100~\cite{manilow2019cutting} dataset. 
To enable long-term generation, we formulate accompaniment generation as a sliding-window inference protocol, where the model continuously inpaints new segments as the window advances.
For real-time operation, we train the model in a \emph{Look-ahead} regime: generating audio for a future time point via outpainting, decoupling the generation timeline from playback so that previously generated audio plays back while the next segment is computed.

In practice, we find that the look-ahead mechanism introduces an inherent trade-off: the deeper the look-ahead, the more the model must generate over unseen future context — particularly challenging for LDMs, whose fixed-length receptive field leaves little observed context to condition on, degrading generation quality. Additionally, the required look-ahead depth is directly related to model's speed — and diffusion models require many denoising steps that makes them notaruisly slow.
To mitigate this, we apply consistency distillation~\cite{song2023consistency, Kim24CTM} to our LDM model achiving considerable speedup and enabling real-time operation with shortened look-ahead step. 

We systematically evaluate our models on the Slakh2100 test set to study the quality-latency trade-off across multiple look-ahead configurations, benchmarking against the baseline StreamMusicGen~\cite{wu2025streaming}. In accordance with the baseline, both of our models perform strongly in the Retrospective regime and degrade gracefully as look-ahead increases — demonstrating the feasibility of the approach while highlighting the challenges that remain.

For live performance, the models are deployed on a Python server communicating via Open Sound Control (OSC)~\cite{WRIGHT_2005}. As a client, we develop a custom MAX/MSP external that supports both local and remote GPU server communication, transmitting audio chunks for inference with minimal latency and tight integration into MAX/MSP's native audio processing system. The external is designed to be model-agnostic, potentially supporting any future accompaniment generation model with minimal setup requirements. On top of this external, we build the Real-Time Accompaniment Patch (RTAP) — a ready-to-use MAX/MSP performance patch handling real-time audio capture, buffering, and playback, bridging the gap between machine learning infrastructure and live performance.

\begin{figure*}[]
  \centering
  \includegraphics[width=0.95\linewidth]{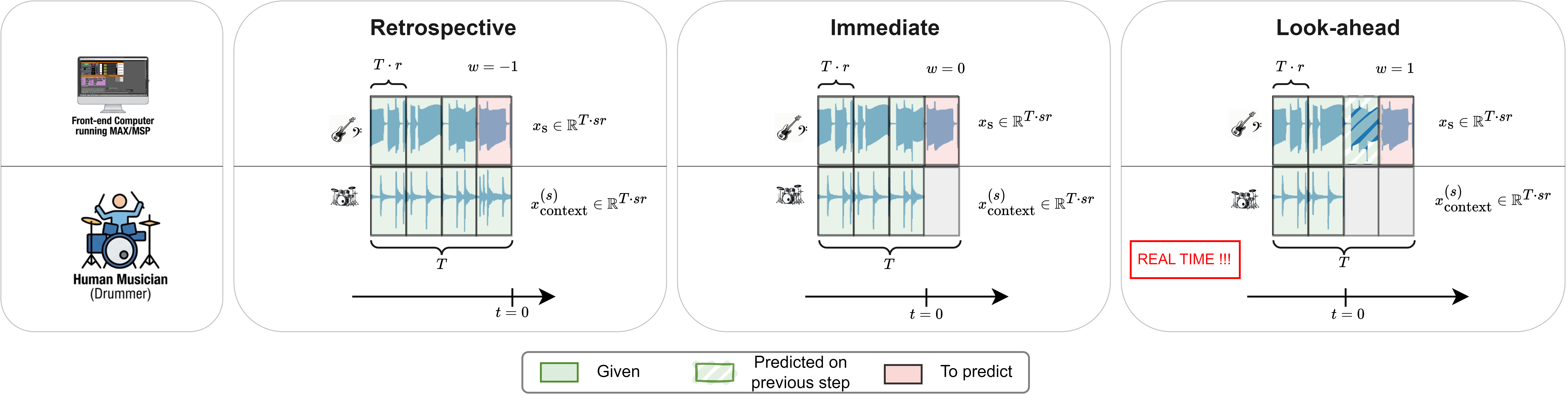}
  \caption{%
    \textbf{Sliding-window protocol for continuous/streaming accompaniment generation.}
    At each step, the window advances by $T \cdot r$, where $r$ controls the step size, and the model predicts a new audio segment conditioned on the available context. Green regions indicate audio that is given (either observed from the performer or generated in previous steps), while red regions denote the segment to be generated. The horizontal axis represents time, with $t = 0$ marking the current playback position. We illustrate three regimes corresponding to look-ahead depth $w \in \{-1, 0, 1\}$: (left, $w < 0$) Retrospective prediction on already-observed content; (middle, $w = 0$) Immediate prediction without look-ahead, insufficient for uninterrupted playback under non-negligible inference latency; (right, $w > 0$) explicit Look-ahead, where audio is generated ahead of playback time, enabling continuous real-time streaming.
    }

  \label{fig:real-time-protocol}
\end{figure*}

Our contributions are as follows. We formalize the look-ahead training and sliding-window inference paradigm for LDM-based real-time accompaniment, with a parameterization we believe will be useful and generalizable to future diffusion-based systems. We develop a model-agnostic MAX/MSP external together with a ready-to-use performance patch, enabling anyone to connect their own generative model to a live performance environment. Finally, to support future research, we release all components: model code with pre-trained checkpoints\footnote{\url{https://github.com/karchkha/musical-accompaniment-ldm}}, the MAX/MSP external and patch\footnote{\url{https://github.com/karchkha/multi_track}}, and a demo page with audio examples\footnote{\url{https://karchkha.github.io/musical-accompaniment-demo/}}.

\section{Related Work}
\label{sec:related_work}

\textbf{Music-to-Music and Accompaniment Generation:}\space\space
A growing body of work addresses generating musical accompaniment conditioned on musical context audio, rather than on text. SingSong~\cite{donahue2023singsong} produces instrumental accompaniment from a vocal recording. StemGen~\cite{parker2024stemgen} trains a non-autoregressive transformer conditioned on a mixture to synthesize a coherent new stem. For bass specifically, Pasini et al.~\cite{pasini2024bass} propose a conditional LDM with timbre control via a reference sample. Diff-A-Riff~\cite{nistal2024diffariff} generates accompaniment conditioned on a user-provided reference, later improved with a diffusion transformer backbone~\cite{nistal2024diffariff2}. Several works target full multi-track generation: MSDM~\cite{mariani2024msdm} unifies generation and source separation in a single waveform diffusion model, JEN-1 Composer~\cite{yao2025jen1composer} presents a unified high-fidelity multi-track framework, and MusicGen-Stem~\cite{rouard2025musicgen} extends autoregressive modeling to multi-stem generation and editing. Our prior work, MT-MusicLDM~\cite{karchkhadze2024multitrack} and MSG-LD~\cite{karchkhadze2025Simultaneous}, operates in the latent diffusion domain for conditional accompaniment generation and joint separation. Xu et al.~\cite{xu2024multisource} improve audio quality with per-source VAEs, and MGE-LDM~\cite{chae2025mgeldm} further advances joint latent diffusion for simultaneous generation and extraction.

A parallel line of work addresses accompaniment generation in the symbolic MIDI domain, including harmonization~\cite{Paiement2006ProbabilisticMH, simon2008mysong}, bass and percussion generation~\cite{lattner2019high, grachten2020bassnet}, and multi-track generation~\cite{MusGan2018, MMM2020, MTMG20, dong2023multitrack}. These works operate on symbolic representations and lie outside our audio-domain real-time focus.

\textbf{Real-time Music Accompaniment Systems:}\space\space
Real-time musical accompaniment by a machine has been a natural long-standing goal in computer music research~\cite{dannenberg1984line, kim2026designspace}. Classical approaches are predominantly rule-based: score-following systems synchronize a pre-authored accompaniment to a live performance by aligning it against a notated score~\cite{dannenberg1984line, raphael2010music, cont2008antescofo}, while others generate responses via hand-crafted heuristics~\cite{lewis2003too} or by recombining phrases drawn from a corpus~\cite{assayag2006omax, nika2012improtek, nika2017improtek}. Recent systems incorporate learned models, though most target symbolic representations such as note counterpoint or chord sequences~\cite{benetatos2020bachduet, wang2022songdriver, jiang2020rl, wu2024adaptive, scarlatos2025realjam}. Magenta RealTime~\cite{team2025live} takes a step toward audio-domain interaction, generating a continuous acoustic stream that responds to user-specified weighted text prompts.

Most closely related to our work is recently published StreamMusicGen~\cite{wu2025streaming}, which proposes a decoder-only Transformer for streaming audio-to-audio accompaniment on the Slakh2100 dataset. To support real-time streaming, it explicitly formulates a look-ahead inference paradigm and systematically studies the trade-off between future context visibility and generation quality across three model variants. While it provides a solid theoretical framework, it does not include a tool for interacting with the model in real time, as our system does. We use this work as our baseline and compare against it using the same evaluation metrics and dataset.

\textbf{Look-ahead and Latency Compensation:}\space\space
A recurring strategy for managing processing latency in real-time systems is to pre-generate future outputs so that a buffer of ready content absorbs computation delay. In classical control, the Smith predictor~\cite{Smith1959} and model predictive control~\cite{Schwenzer2021} address this by forecasting system behavior over a finite horizon and scheduling commands in advance. In robot learning, action chunking~\cite{black2025real} generates a short sequence of future actions per inference step, providing an execution buffer while the next prediction is computed. Our look-ahead conditioning follows the same principle: the model generates audio for a future window before it is needed for playback, using that lead time to hide diffusion inference latency.

\section{Method}
\label{sec:method}


Fig.~\ref{fig:system_overview} gives an overview of the system we propose for real-time interactive musical accompaniment, in which a human performer plays live while an LDM generates matching instrumental parts. Real-time responsiveness is achieved through a client–server architecture: the server runs the inference-heavy LDM in a Python backend, while the client—a MAX/MSP patch built around a custom external object—interfaces directly with the musician’s setup. Both components are designed to operate under the practical constraints of buffering, latency, and streaming.

The remainder of this section is organized as follows. Section~\ref{sec:streaming} formalizes the sliding-window streaming inference protocol and defines the Look-ahead regime. Section~\ref{sec:models} describes the LDM backbone, its training objective, and the consistency distillation procedure used to accelerate inference. Section~\ref{sec:maxmsp} details the MAX/MSP client implementation that connects the generative model to a live performance environment.

The following notation is used throughout. We denote a musical mixture as \( x_{\text{mix}} = \sum_{s=1}^{S} x_s \), where \( x_s \in \mathbb{R}^{T \cdot sr} \) are \( S \) mono stems, \( T \) is duration in seconds, and \( sr \) is the sample rate. The task of accompaniment generation is to synthesize a target stem \( \hat{x}_s \) conditioned on the remaining context \( x_{\text{context}}^{(s)} = \sum_{i \neq s} x_i \), which we write compactly as \( \hat{x}_s = f(x_{\text{context}}^{(s)}) \).

\begin{figure*}[ht]
  \centering
  \includegraphics[width=\linewidth]{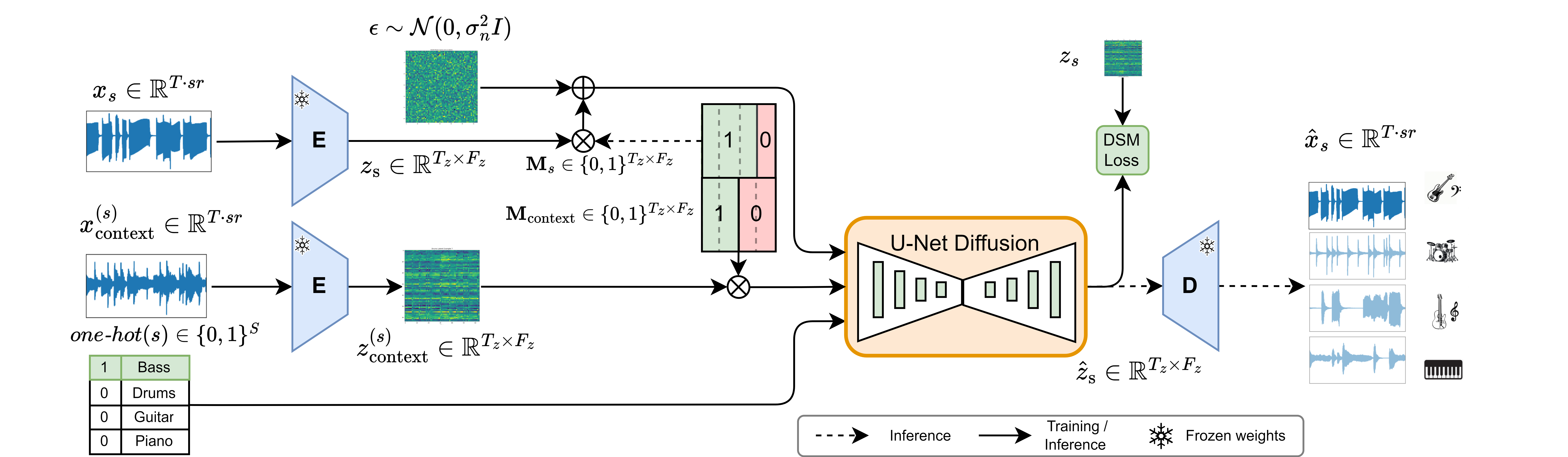}
  \caption{
        \textbf{Latent diffusion model for accompaniment generation.}
        The input mixture is encoded into a latent representation by encoder $E$, then concatenated with a noise sample and a stem-selection one-hot vector. A SongUNet-based diffusion model iteratively denoises the target stem latent, conditioned on the mixture context. Decoder $D$ reconstructs the waveform. For streaming operation, the model is trained and inferred in inpainting mode, where the portion of the target latent corresponding to already-generated content is kept fixed while only the future segment is denoised.
        }   
  \label{fig:latent_diffusion_model}
\end{figure*}

\subsection{Sliding-Window Streaming Protocol}
\label{sec:streaming}

We formulate accompaniment as a sliding-window protocol operating over a fixed-length audio receptive field of duration $T$. We define the step size as $T \cdot r$, where $r \in (0, 1]$ controls the fraction of the window advanced at each step: smaller $r$ yields finer-grained, more frequent updates; larger $r$ yields coarser, less frequent ones. At each discrete step $t$, rather than generating the full accompaniment from scratch, the model conditions on the current mixture context $x_{\text{context, t}}^{(s)}$ together with the previously generated segment $\hat{x}_{s,t-1}$ shifted forward in time by $r \cdot T$. Denoting this time-shift operator as $\mathcal{S}_{rT}$, the generation at step $t$ is:
\begin{equation}\label{eq:sliding_window}
  \hat{x}_{s,t} = f(x_{\text{context, t}}^{(s)} \mid \mathcal{S}_{rT}(\hat{x}_{s,t-1})).
\end{equation}
The shift places the previous prediction into the past portion of the new window, leaving only the final $r \cdot T$ segment unknown. The model thus inpaints solely this novel region, while the overlapping $(1-r) \cdot T$ portion is kept fixed. The window then advances by $T \cdot r$ and the process repeats, enabling continuous generation as previously generated content is fed back with newly observed context.

We additionally define two system-level latency parameters: $d$, the time required by the generative model to produce a single prediction step, and $\delta$, the audio buffer processing time of the host environment (e.g., a DAW or, in our case, MAX/MSP). The interaction between $d$ and $\delta$ determines whether uninterrupted real-time playback can be achieved.

Fig.~\ref{fig:real-time-protocol} illustrates the sliding-window protocol (green: observed or generated audio; red: current target segment). We define the look-ahead depth $w \in \mathbb{Z}$ as the number of step-sized intervals $T \cdot r$ between the current playback position and the start of the predicted window (always the last $T \cdot r$ portion of the target). As illustrated, $w$ determines where the current playback position $t{=}0$ falls within the receptive field of duration $T$, defining how much context is available and whether prediction is performed via inpainting (target lies within available context) or outpainting (target extends beyond it). We distinguish three regimes based on the sign of $w$:
\begin{itemize}
    \item \textbf{Retrospective ($w < 0$):} The predicted windows lie entirely in the past, thus a full musical context is available. This is the classic accompaniment generation setting. In our work it serves as a reference point and an upper bound on achievable quality and coherence.

    \item \textbf{Immediate ($w = 0$):} The predicted window starts at the current playback position. To support real-time operation, this setting requires $d \leq \delta$. However, unlike traditional real-time digital audio processing, diffusion models carry non-negligible inference latency $d \gg \delta$, making real-time unachievable in this setting.

    \item \textbf{Look-ahead ($w > 0$):} The predicted window starts $w$ steps ahead of the current playback position, creating a temporal buffer (illustrated by the striped green region in Fig.~\ref{fig:real-time-protocol}) that allows previously generated audio to play back while the next step is computed. This decouples generation from playback, enabling uninterrupted real-time operation even when $d > \delta$. 

\end{itemize}

\vspace{1pt}
While, for real-time performance, $w$ can in principle be any positive integer, larger values reduce available context: only $(1-(w+1) \cdot r) \cdot T$ of the receptive field contains known audio, so even at $w = 1$ the system effectively operates with a two-step generation horizon. In our setting, we set $w = 1$ under the constraint $T \cdot r \geq d$, ensuring inference completes before playback reaches the predicted segment. The choice of $r$ governs a fundamental trade-off: smaller $r$ shortens the prediction horizon and eases the task but tightens the latency budget, while larger $r$ relaxes that constraint, observed empirically to reduce musical coherence — see Section~\ref{sec:results}. 
Note that this formulation is presented in the audio time domain; its extension to the latent space, where $r$ is further constrained by the latent grid resolution, is detailed in Section~\ref{sec:lookahead}.

\subsection{Accompaniment Generation Models}
\label{sec:models}
\vspace{5pt}

For the accompaniment generation task, we develop a generative model using a latent diffusion (LDM) paradigm that synthesizes a target instrument stem conditioned on a latent representation of the mixture context $z^{(s)}_{\text{context}}$ and a one-hot instrument label $s$. We first describe the audio compression into latent vectors, then the LDM architecture, then the consistency distillation (CD) procedure used to reduce inference latency, and finally the training adaptations that enable the model to operate within the streaming protocol of Section~\ref{sec:streaming}.

\subsubsection{Audio Compression with Music2Latent}

We operate our generative models in a compressed latent space achieved using the pre-trained Music2Latent~\cite{Pasini2024Music2Latent} autoencoder. Music2Latent is a consistency-based convolutional autoencoder designed for efficient, high-fidelity audio compression. As depicted in Fig.~\ref{fig:latent_diffusion_model}, the encoder $E$ maps a mono audio signal $x$ into a 2D latent representation $z \in \mathbb{R}^{T_z \times F_z}$, where $T_z$ and $F_z$ denote the temporal and frequency dimensions, respectively. The decoder $D$ reconstructs the waveform $x$ from $z$. Given this mapping $x \leftrightarrow z$, the generation problem is formulated as modeling $q(z)$ instead of $q(x)$, allowing the diffusion model to operate in a stable, low-dimensional latent space.

\subsubsection{Latent Diffusion Model}

As depicted in Fig.~\ref{fig:latent_diffusion_model}, we employ a latent diffusion model (LDM)~\cite{liu2023audioldm, chen2024musicldm} operating on the compressed latent space $z$. The generative backbone $g_\phi$ is a U-Net architecture based on the SongUNet~\cite{song2021scorebased, karras2022elucidating}, adapted to operate on the 2D latent representations produced by Music2Latent. The network uses positional timestep embeddings and a standard DDPM++ encoder–decoder structure with resampling filters.

For the diffusion part, we adopt the denoising score-matching (DSM)~\cite{song2019generativemodeling, song2021scorebased} paradigm, where the model learns the score function $\nabla_z \log p(z)$—the gradient of the log-density, which points toward cleaner, higher-likelihood samples. Given a data point $z_0 \sim p(z_0)$, noise-corrupted samples are constructed as $z_n = z_0 + \sigma_n \epsilon$, where $\epsilon \sim \mathcal{N}(0, I)$ and $\sigma_n$ controls the noise level. By following the estimated score field iteratively, the trained model denoises from pure noise back toward the clean data distribution.

The model is trained with denoising score matching and sampled via ODE integration, following the EDM framework~\cite{karras2022elucidating}. During training, the noise level $\sigma$ is sampled from a log-normal distribution $\ln(\sigma) \sim \mathcal{N}(P_\text{mean}, P_\text{std}^2)$, and the model $g_{\phi}$ is optimized using the DSM objective:
\begin{equation}\label{eq:dsm_loss_ours}
    \scalebox{0.95}{$
    \mathcal{L}_{\text{DSM}}(\phi) =
    \mathbb{E}_{s, z_{\text{context}}, n}
    \left\|
        z_s -
        g_{\phi}\bigl(z_s + \sigma_n \epsilon,\; s,\; \sigma_n,\; z^{(s)}_{\text{context}}\bigr)
    \right\|_2^2.
    $}
\end{equation}

Inference with $g_{\phi}$ is performed as an iterative numerical integration of an ODE over $N$ discrete steps. At each step, $g_{\phi}$ denoises the target stem $\hat{z}_{s,n}$ conditioned on the context signal $z^{(s)}_{\text{context}}$, the instrument identity $s$, and noise level $\sigma_n$. The noise schedule is defined as a mapping $\sigma : \{1, \dots, N\} \rightarrow [\sigma_{\min}, \sigma_{\max}]$, and the process is initialized from pure noise $\hat{z}_{s,N} \sim \mathcal{N}(0, \sigma_{\max}^2 I)$, then progressively denoised according to the update rule:
\begin{equation}\label{eq:ode_solver}
    \hat{z}_{s, n-1} = \texttt{Solver}\bigl(\hat{z}_{s, n}, s, \sigma_n, z^{(s)}_{\text{context}}; g_{\phi}\bigr),
\end{equation}
where $\texttt{Solver}$ denotes the DPM2~\cite{lu2022dpmsolver}—a fast second-order ODE solver achieving high-quality generation in very few steps—with the Karras noise schedule~\cite{karras2022elucidating}.

\begin{figure}[t]
  \centering
  \includegraphics[width=\linewidth]{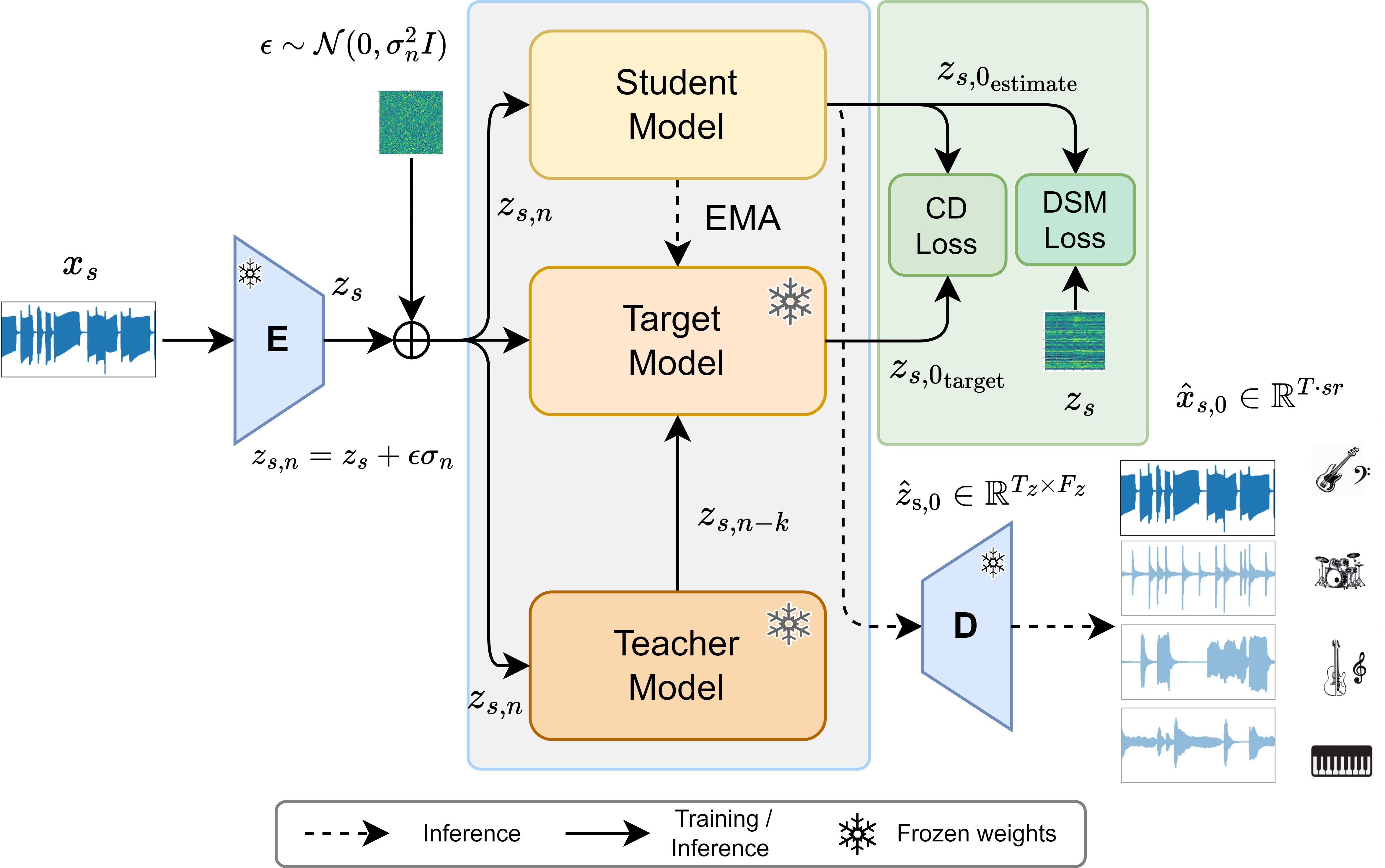}
  \caption{%
    \textbf{Consistency distillation.}
    A pretrained diffusion model (teacher) generates denoised targets by applying $k$ ODE steps along the diffusion trajectory. A student model is trained to directly predict these targets from noisy inputs, enforcing consistency across noise levels. An EMA of the student parameters provides stable training targets. The training objective combines the CD loss with a DSM loss. Latent context conditioning and instrument label inputs are omitted from the figure for clarity.}
  \label{fig:CD}
\end{figure}

\subsubsection{Consistency Distillation}

While the EDM framework achieves good generation quality in a relatively small number of steps, the iterative denoising process still introduces latency that challenges real-time operation. To further reduce this bottleneck, we apply consistency distillation (CD)~\cite{song2023consistency, Kim24CTM}. The key idea is to train a student model $g_{\omega}$ to map any point along the diffusion trajectory directly to the clean data estimate, bypassing the need for full iterative denoising and enabling generation in one or two steps.

We follow the CTM~\cite{Kim24CTM} formulation exactly. As illustrated in Fig.~\ref{fig:CD}, the student $g_{\omega}$ is architecturally identical to the teacher $g_{\phi}$ and is trained to produce outputs consistent across noise levels along the probability flow ODE. Teacher targets are generated by applying $k$ ODE solver steps from noise level $\sigma_n$ to a lower level $\sigma_{n-k}$:
\begin{equation}
    \hat{z}_{s, n-k}^{\text{dif}} =
    \texttt{Solver}_k\!\left(
    z_s + \sigma_n \epsilon,\,
    s,\,
    \sigma_n,\,
    z^{(s)}_{\text{context}};\,
    g_{\phi}
    \right),
\end{equation}
where $k \in [1, n]$ denotes the number of iterative ODE solver steps. The student is then trained to match these targets via the CD loss~\cite{Kim24CTM}:
\begin{equation}
\begin{aligned}
\mathcal{L}_{\text{CD}}(\omega)
&= \mathbb{E}_{n,k} \Big\|
\underbrace{
g_{\texttt{sg}(\omega)}\!\left(
\hat{z}_{s, n-k}^{\text{dif}},
s, \sigma_{n-k}, z^{(s)}_{\text{context}}
\right)
}_{\textit{target}} \\
&\quad -
\underbrace{
g_{\omega}\!\left(
z_s + \sigma_n \epsilon,
s, \sigma_n, z^{(s)}_{\text{context}}
\right)
}_{\textit{estimate}}
\Big\|_2^2,
\end{aligned}
\end{equation}
where $\texttt{sg}(\omega)$ is a stop-gradient EMA of the student parameters: $\texttt{sg}(\omega) \leftarrow \texttt{stopgrad}(\mu\,\texttt{sg}(\omega) + (1-\mu)\,\omega)$.

Farther following CTM, we augment the CD loss with the addition of DSM loss from Eq.~\eqref{eq:dsm_loss_ours} to provide direct data supervision:
\begin{equation} \label{eq:our_final}
\mathcal{L}({\omega}) = \mathcal{L}_{\text{CD}}({\omega}) + \lambda_{\text{DSM}}\mathcal{L}_{\text{DSM}}(\omega),
\end{equation}
where $\lambda_{\text{DSM}}$ balances the two terms. 

For inference, we similarly follow CTM and use the multistep consistency sampler.

\subsubsection{Look-Ahead Training Adaptation via Latent Masking}
\label{sec:lookahead}

In the Look-ahead regime ($w = 1$), the model must generate future audio segments with part of the musical context unavailable. Since the sliding-window protocol (see Section~\ref{sec:streaming}) was formulated in the audio time domain but the LDM operates in the compressed latent space of Music2Latent, we describe here how this regime is realized in the latent space via a masked conditioning strategy.

The key training adaptation for the LDM model to support all three regimes is a masked context conditioning strategy (depicted in Fig.~\ref{fig:latent_diffusion_model}). Since the generative model has a fixed receptive field of length $T$, the position of the current playback moment within the window depends on $w$ and $r$. As illustrated in Fig.~\ref{fig:real-time-protocol}, in the Retrospective regime ($w = -1$), the current time lies at the right edge, the predicted window starts in the past, and no context is missing. In the Immediate regime ($w = 0$), the predicted step begins exactly at the current time, so the last $T \cdot r$ of the window is future and unavailable, shifting the current time to $T \cdot r$ from the right edge. In the Look-ahead regime ($w = 1$), the current time shifts further left to accommodate one additional future step, placing it $2 \cdot T \cdot r$ from the right edge. Thus, across all regimes, current time sits at $(w+1) \cdot T \cdot r$ seconds from the right edge, and the corresponding portion of context is unavailable and must be masked. Operating in the latent space with temporal resolution $T_z$ and frequency resolution $F_z$, we define a binary context mask $\mathbf{M}_{\text{context}} \in \{0,1\}^{T_z \times F_z}$ as:
\begin{equation}\label{eq:context_mask}
\mathbf{M}_{\text{context}}[t,f] =
\begin{cases}
1, & \text{if } t < T_z - T_z \cdot r \cdot (w+1), \\
0, & \text{otherwise.}
\end{cases}
\end{equation}
where $r \in (0,1]$ is the step ratio, which must be chosen such that $T_z \cdot r \in \mathbb{N}$—i.e., the step maps to a whole number of latent frames, as fractional frame boundaries cannot be masked. The masked context is obtained as $z_{\text{context}} \odot \mathbf{M}_{\text{context}}$, zeroing out the future portion. To expose the model to varying degrees of missing context, $r$ is randomly sampled during training, enabling robust generation across different look-ahead configurations at inference time.

At inference, the context latent is masked identically to training using $\mathbf{M}_{\text{context}}$. The target latent, however, is masked differently. As established in Eq.~\ref{eq:sliding_window}, the segment to be generated is always the last $r \cdot T$ of the target stem; the remainder was already generated in the previous step, placed into the window via the $\mathcal{S}_{rT}$ shift, and kept fixed while last part is inpainted. Thus, only this final segment needs to be synthesized, giving the target mask:
\begin{equation}\label{eq:target_mask}
\mathbf{M}_{s}[t,f] =
\begin{cases}
1, & \text{if } t < T_z - T_z \cdot r, \\
0, & \text{otherwise.}
\end{cases}
\end{equation}
In the inference process, the unmasked region ($\mathbf{M}_{s} = 1$) is filled with previously generated content from the prior step and kept fixed; the masked region ($\mathbf{M}_{s} = 0$) is initialized with Gaussian noise and iteratively denoised by the inpainting sampler. This asymmetric masking preserves temporal continuity across steps while ensuring only the truly new segment is generated at each step.

\subsection{MAX/MSP Integration}
\label{sec:maxmsp}
Max/MSP~\cite{cycling74max} is a visual programming environment for music and multimedia developed by Cycling '74. Max offers the flexibility to build complex systems with custom logic and UI. It is widely adopted among musicians and integrates natively with Ableton Live (via Max for Live), hardware controllers, and OSC-based workflows, making it a natural choice for performance-oriented tools. We implement the client-side component of the system in Max/MSP, leveraging its real-time audio buffering, temporal scheduling, and extensible UI capabilities. The performer interacts exclusively with the Max/MSP patch to configure instrument roles and trigger generation, while all computationally intensive inference is handled by the Python server, either locally or remotely. The design of the system is intentionally made agnostic to the neural network architecture: with minimal changes on the server and Max side, the same patch and external can be connected to any generative model.

\subsubsection{The \texttt{multi\_track} Max/MSP External}

We developed \texttt{multi\_track}, a custom Max/MSP external written in C++. While Max/MSP natively provides all the building blocks for such a server communication with OSC messages, implementing this at the C++ level bypasses Max’s control-rate signal constraints, enabling significantly faster performance than would be achievable in a native patch. The external is integrated with the Max/MSP environment and acts as the sole interface between the Max/MSP environment and the Python inference server, managing the full lifecycle of audio data: reading from a shared multichannel \texttt{buffer\~{}}, transmitting context to the server over UDP using the OSC protocol, and writing predictions back into the \texttt{buffer\~{}} as they arrive.

The external takes as arguments the name of the \texttt{buffer\~{}} to bind and the channel names for each instrument stem (bass, drums, guitar, piano), which must correspond to the channel layout of the buffer. The \texttt{buffer\~{}} serves as the shared memory space for the entire real-time loop: incoming audio from the performer is written into the buffer by standard Max objects, while the external reads context from it and writes generated audio back into it on a per-prediction basis. The external additionally takes a \texttt{predict\_instruments~<array>} message, where array specifies which stem the server should generate (i.e., \texttt{predict\_instruments 1 0 0 0} predicts the first stem — bass in our case). All non-predicted channels are summed into a single mixture and sent to the server under \texttt{/context}; an alternative mode sends each stem independently, left for future versatility.

Each prediction cycle is triggered by a \texttt{predict~<curr>} message, where \texttt{curr} is the position of the most recently crossed step boundary — i.e., a multiple of $r \cdot T$ samples (e.g., for $r{=}0.25$ and $T{=}6\,\text{s}$: $0, 1.5, 3, 4.5, \ldots$). The external directly operates the streaming parameters $T$, $r$, and $w$ defined in Section~\ref{sec:streaming}. Context and generated audio are exchanged in a windowed, stepped manner. Context is always read and sent from the interval $[\texttt{curr} - r \cdot T,\, \texttt{curr}]$, while the generated output is written to the corresponding stem buffer at $[\texttt{curr} + w \cdot r \cdot T,\, \texttt{curr} + (w+1) \cdot r \cdot T]$, directly implementing our Retrospective, Immediate, and Look-ahead regimes.

\begin{figure}[t]
  \centering
  \includegraphics[width=\columnwidth]{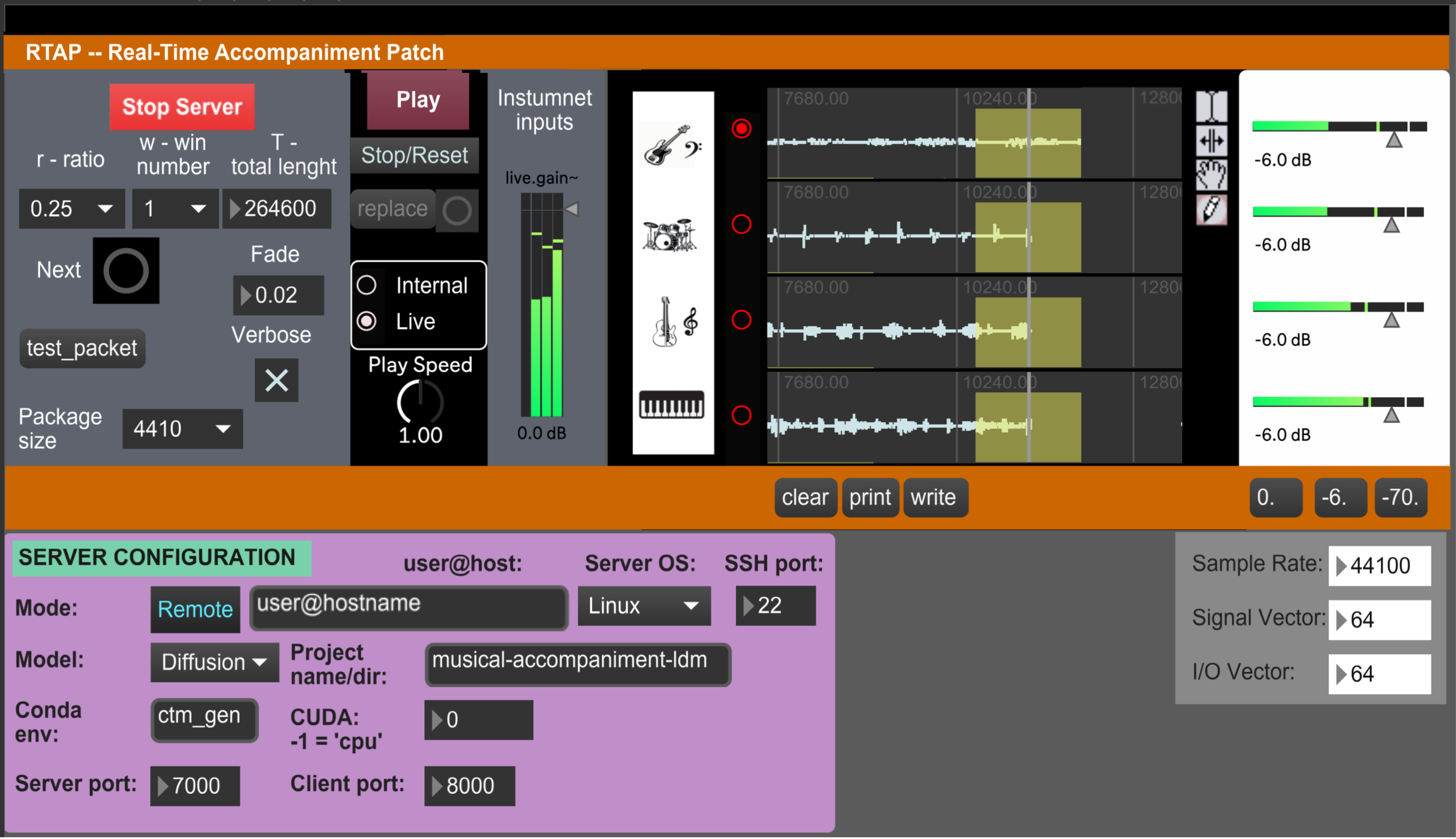}
  \caption{\textbf{The RTAP MAX/MSP patch.} The main panel controls streaming parameters ($r$, $w$, $T$), instrument selection (bass, drums, guitar, piano), and playback (\texttt{Play}/\texttt{Stop/Reset}). \texttt{Fade} crossfades between chunks to suppress boundary clicks; \texttt{Package size} sets OSC chunk size. The \texttt{Internal}/\texttt{Live} toggle selects between buffer playback and live microphone input; \texttt{Internal} mode adds a \texttt{Source} file selector (inactive here). Waveform displays and output meters provide real-time monitoring. The lower-left panel configures the server: inference mode (Diffusion or CD), credentials (host, ports, conda environment, project path), and CUDA device. In the screenshot, the patch runs in remote \texttt{Live} mode: microphone inputs provide context for drums, guitar, and piano; the first buffer shows the predicted bass stem; the highlighted window is the current step — live-recorded for the other instruments and bass-predicted on the previous step — with the next prediction already in flight.}
  \label{fig:max_patch}
\end{figure}

Audio context is transmitted to the server as a stream of fixed-size UDP packets, whose size is controlled by the \texttt{packet\_size} argument of the external. Each packet carries an OSC-formatted message containing a step identifier, a chunk index, the total expected chunk count, and a block of floating-point audio samples. The step identifier allows the server to detect and discard stale responses from a previous prediction cycle that arrive after a new one has begun, while the chunk index allows the server to gather and place audio samples at the corresponding positions in the running tensor, and to self-trigger inference as soon as all chunks arrive. Predicted audio is returned from the server as the same chunk-based OSC stream. The external maintains a persistent listener thread that receives incoming packets and writes each chunk directly into the \texttt{buffer\~{}} at the appropriate positions as they arrive. A \texttt{fade} argument passed to the external is forwarded to the server, which returns that many extra samples prepended before the nominal write window; the external applies a linear fade-in over those samples to suppress boundary clicks at the junction between generated audio on consecutive steps.

The external also manages the server lifecycle: a \texttt{set\_command} message passes the full shell command used to launch the Python server, which the external spawns directly and monitors from within Max/MSP. The server may run locally or on a remote GPU machine accessed over SSH. UDP port numbers for both communication directions are configurable at runtime, and the server and client IP addresses are determined automatically and injected into the launch command before the server process starts.

\subsubsection{Python Inference Server}

On the server side, the Music2Latent encoder/decoder and the diffusion or consistency model are loaded onto GPU (or MPS on Apple Silicon) and held ready between inference calls. Additionally, the server maintains two running buffers — context audio and generated audio latent vectors— that are shifted left by $r \cdot T_z$ after each cycle, keeping them aligned with the sliding window without requiring explicit synchronization with the Max side. Incoming audio chunks form MAX are received over UDP and gathered; once all chunks of a batch arrive, inference is triggered automatically. At inference time, the server encodes context audio into the latent space via the Music2Latent encoder, zeros out the prediction region using $\mathbf{M}_{\text{context}}$ (Eq.~\ref{eq:context_mask}), and runs the diffusion or consistency distillation inpainting step. The predicted latent is decoded back to audio and streamed immediately to Max/MSP as chunked OSC messages.

\subsubsection{RTAP: Real-Time Accompaniment Patch}

The \texttt{multi\_track} external is wrapped in the RTAP (Real-Time Accompaniment Patch), a dedicated Max/MSP patch that provides a complete performance interface: audio capture and playback, parameter control, server management, and real-time monitoring. While the current configuration specifically targets our diffusion model with context-to-accompaniment generation with four stems, the patch is designed to be versatile: the number of instruments, sample rate, and receptive field duration can all be reconfigured with minimal changes to the external arguments and server settings, and it should work with any generative model that can be adapted to our streaming protocol.

As shown in Fig.~\ref{fig:max_patch}, the main panel of the patch exposes the streaming parameters $T$, $r$, and $w$ as numeric controls, defining the model's receptive field length, step size, and Look-ahead regime. The instrument selector determines which stem is generated. Audio input can come from a live microphone (\texttt{Live} mode) or a pre-loaded audio file (\texttt{Internal} mode), selectable via a toggle. The \texttt{Fade} parameter controls the crossfade length as a ratio of the sample rate between consecutive generated chunks, and \texttt{packet\_size} sets the OSC chunk size for network transmission. The server can be started and stopped via the \texttt{Start/Stop Server} button. A \texttt{Next} button manually triggers a prediction cycle without requiring playback to cross a step boundary, useful for testing or on-demand generation. A \texttt{Test packet} button verifies the OSC connection, and a \texttt{Verbose} toggle prints detailed OSC message logs on both the Max and server sides. A \texttt{Clean} button resets all buffers and server-side tensors; \texttt{Print} saves images of the current audio and latent vectors to disk for debugging; \texttt{Write} dumps the recorded and generated audio as a multichannel file for offline review. Waveform displays and output level meters provide continuous visual feedback during performance.

The lower purple server configuration panel allows the performer to specify the inference mode (Diffusion or CD), server address and SSH credentials for remote operation, conda environment, project path, and CUDA device.

\begin{figure*}[!t]
  \centering
  \includegraphics[width=\textwidth]{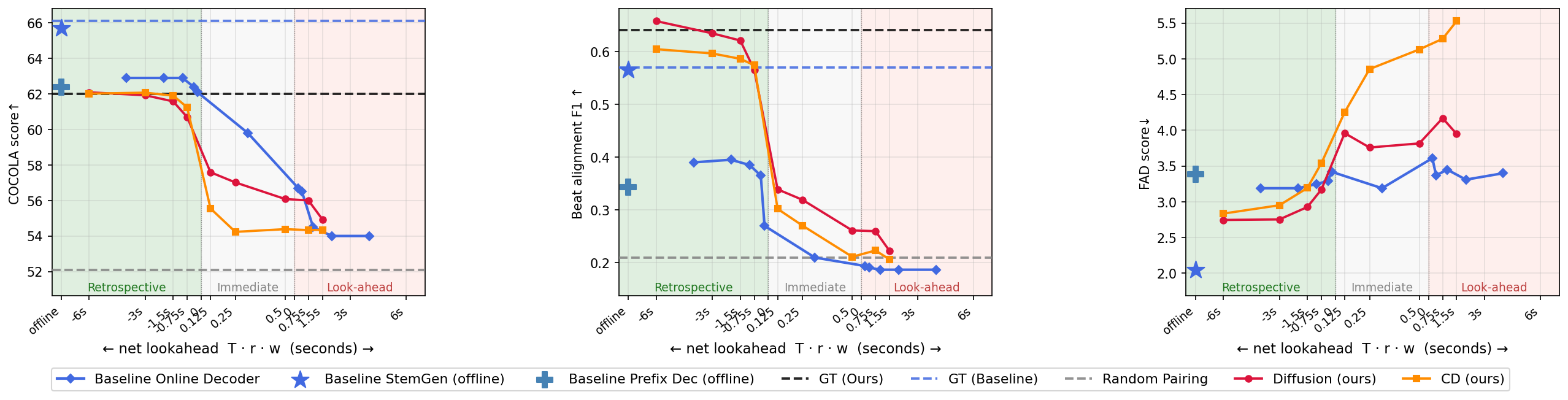}
  \caption{Comparison of our diffusion model and consistency distillation (CD) model against paper baselines across three evaluation metrics. The x-axis represents the net look-ahead $T \cdot r \cdot w$ (seconds), partitioned into three zones: \textit{Retrospective} ($w{=}{-1}$, green), \textit{Immediate} ($w{=}0$, gray), and \textit{Look-ahead} ($w{=}1$, salmon). Points in the Immediate zone are spread by step size $r$.}
  \label{fig:sweep_analysis}
\end{figure*}

\section{Experimental Setup}
\label{sec:experiment_setup}

This section covers the experimental setup for both components of the system. First, we describe the generative model setup: dataset, model architecture, training procedure, baselines, and evaluation metrics used to assess accompaniment generation quality across different streaming configurations. Second, we describe the RTAP system configuration used for real-time latency evaluation, including the hardware, network, and OSC transmission parameters of the deployed client--server system. For full implementation details, training configurations, and model checkpoints, we refer the reader to the code repositories linked above.

\subsection{Generative Model Setup}

\subsubsection{Dataset}

We trained our models on the Slakh2100 dataset~\cite{manilow2019cutting}, a synthesized multi-track MIDI dataset rendered at 44.1 kHz sampling rate. The dataset was split into training, validation, and test subsets. We focused on four instrumental stems: bass, drums, guitar, and piano.

We extracted 6-second segments (264,600 samples at 44.1\,kHz), with the extraction window randomly shifted by up to $\pm\frac{1}{2}$ the segment length for augmentation. For each sample, one target stem $x_s$ and its one-hot class label $s$ were randomly selected, and the context $x^{(s)}_{\text{context}}$ was formed by summing all remaining stems.

\subsubsection{Model Settings}

The pre-trained Music2Latent encoder maps a mono audio segment $x$ of shape $1 \times 264{,}600$ to latent codes $z$ of shape $1 \times 64 \times 64$, corresponding to a temporal compression factor of $264{,}600 / 64 \approx 4{,}134$, with the remaining compression realized through projection into a compact frequency dimension of 64 bins.
The decoder reconstructs audio waveforms from these latent representations. During training, both encoder and decoder remain frozen, allowing the diffusion model to learn in a stable, pre-defined latent space.

The diffusion model $g_{\phi}$ is a U-Net operating on $64 \times 64$ latent images. The encoder progressively downsamples the input from $64 \times 64$ down to $8 \times 8$ across four resolution levels ($64, 32, 16, 8$), with a 2$\times$ spatial downsampling between each level. The channel width is 256 with multipliers $[1, 2, 2, 2]$, and each level contains four residual blocks, yielding approximately 257M parameters. The decoder mirrors this structure with symmetric 2$\times$ upsampling and skip connections from the encoder at each resolution. Self-attention is applied at three scales — $8 \times 8$, $16 \times 16$, and $32 \times 32$ — in both encoder and decoder. The model is conditioned on a 4-dimensional one-hot instrument label $s$ and on the context mixture via channel-concatenation of the latent encoding of $z^{(s)}_{\text{context}}$ with the noisy target latent $z_n$ at the network input, resulting in two input channels and one output channel. Timestep information is injected via positional embeddings with embedding dimension multiplier 4, following the DDPM++ variant of the EDM framework~\cite{karras2022elucidating}. Dropout of 0.10 is applied to intermediate activations.

The consistency model $g_{\omega}$ shares the identical U-Net architecture as $g_{\phi}$. As described in Sec.~\ref{sec:method}, CD involves three model instances: the frozen teacher $g_{\phi}$, the student $g_{\omega}$ being optimized, and a target model $g_{\texttt{sg}(\omega)}$ which is an EMA of the student with a fixed decay rate $\mu = 0.999$.

\subsubsection{Training and Inference Details}

The diffusion model $g_{\phi}$ is trained following the EDM framework~\cite{karras2022elucidating}, with noise levels sampled from a log-normal distribution ($P_{\text{mean}} = -1.2$, $P_{\text{std}} = 1.2$) and data variance $\sigma_{\text{data}} = 1.0$. It is optimized with Adam ($\beta_1 = 0.9$, $\beta_2 = 0.99$) at a learning rate of $10^{-4}$, with a batch size of 64 across 2 NVIDIA RTX A6000 (48 GB) GPUs for 250 epochs. To support the sliding-window streaming protocol described in Sec.~\ref{sec:streaming}, inpainting masks with ratios $r \in \{0, 0.125, 0.25\}$ are applied randomly during training with window offset $w \in \{-1, 0, 1\}$, enabling the model to generate under varying degrees of future context visibility. For inference, we adopt the Karras noise schedule with $\sigma_{\min} = 10^{-4}$, $\sigma_{\max} = 50.0$, and $\rho = 9.0$, and sample using the DPM-2 sampler with $\rho = 1.0$ and 2 resamples per step. We found 5 denoising steps to yield the best results, totalling 10 network forward passes per generation.

The consistency distillation student network $g_\omega$ is initialized from the pre-trained $g_\phi$ and optimized with RAdam ($\beta_1 = 0.9$, $\beta_2 = 0.999$) at a learning rate of $10^{-5}$ with 1 epoch of linear warmup, a batch size of 32 across 2 NVIDIA RTX A6000 GPUs, for up to 50 epochs. During training, the teacher $g_\phi$ generates Heun-solver targets across 18 fixed noise scales, with the number of solver steps drawn uniformly at random up to 17. The same masks ($r \in \{0, 0.125, 0.25\}$, $w \in \{-1, 0, 1\}$) used for $g_\phi$ are also applied during CD training. The total loss follows Eq.~\eqref{eq:our_final} with $\lambda_{\text{DSM}} = 0.7$. At inference, $g_\omega$ generates accompaniment in 1 or 2 steps using the multistep consistency sampler.

\subsubsection{Baselines}
\label{sec:baselines}

We compare our models against three variants from StreamMusicGen~\cite{wu2025streaming}: (1) the Online Decoder, a streaming decoder-only Transformer that generates one chunk at a time and supports varying degrees of future context visibility — analogous to our $w \in \{-1, 0, 1\}$ regimes; (2) the Prefix Decoder, an offline variant with full input context available, corresponding to the Retrospective setting; and (3) StemGen~\cite{parker2024stemgen}, an offline masked language model operating with full context (also our Retrospective setting), serving as an upper bound. These models operate on RVQ tokens from the Descript Audio Codec at 32 kHz, are trained on all instrument categories in Slakh2100, and are evaluated under a future visibility / chunk-size ($t_f$, $k$) paradigm. Our models differ in generative architecture (latent diffusion vs.\ autoregressive transformer), audio representation (Music2Latent latent codes at 44.1 kHz vs.\ RVQ tokens at 32 kHz), instrument set (4 stems vs.\ all Slakh categories), and streaming formulation ($r$, $w$ vs.\ $t_f$, $k$). Consequently, results are not directly numerically comparable, but the comparison provides a useful indication of relative performance across the shared evaluation metrics. In terms of model size, the StreamMusicGen Online and Prefix Decoders contain approximately 294M parameters each, while StemGen uses 348M. This makes the baselines and our models (257M) broadly comparable in scale.

\subsubsection{Evaluation Metrics}

To enable direct comparison with our baseline, we adopt the same evaluation paradigm as StreamMusicGen~\cite{wu2025streaming} and evaluate generated accompaniment across three complementary objective metrics covering musical coherence, rhythmic alignment, and audio quality.

To measure overall coherence, we use the COCOLA~\cite{ciranni2025cocola} score. COCOLA is a self-supervised contrastive model trained to score harmonic and rhythmic coherence between a mixture and a stem. Higher scores indicate greater musical coherence between the generated stem and the input mixture. COCOLA can measure harmonic and rhythmic coherence separately, but for simplicity and consistency with our baselines we report the overall COCOLA score, which is a weighted combination of both.

For rhythmic alignment, we use the Beat Alignment $F_1$ score. Beat positions are estimated in both the input mixture and the generated stem using a beat tracker~\cite{foscarin2024beat} powered by Madmom~\cite{madmom}, and the $F_1$ score between the two sets of beat times is computed. A higher score indicates tighter rhythmic alignment.

For general audio quality, we use Fréchet Audio Distance (FAD)~\cite{kilgour2018fr}, which assesses audio quality by comparing the distribution of VGGish embeddings of generated audio against a reference distribution from the test split. Lower values indicate higher perceptual audio quality.

\begin{table*}[!t]
\centering
\caption{Mean per-stage timings (ms) at $r{=}0.25$ across four deployment configurations. OSC chunk size is 4,410 samples, giving 15 chunks per send. RT: real-time constraint $d < T{\cdot}r = 1500$\,ms satisfied ($\checkmark$) or not ($\times$).}
\vspace{1mm}
\label{tab:latency}
\resizebox{0.9\textwidth}{!}{%
\begin{tabular}{lcccccccc}
\hline
 & \multicolumn{2}{c}{Config (1)} & \multicolumn{2}{c}{Config (2)} & \multicolumn{2}{c}{Config (3)} & \multicolumn{2}{c}{Config (4)} \\

Client  & \multicolumn{2}{c}{Win 10, SD} & \multicolumn{2}{c}{Win 10, SD} & \multicolumn{2}{c}{Mac M2, local} & \multicolumn{2}{c}{Win 10, SD} \\
Server  & \multicolumn{2}{c}{Win 10, RTX 2070, local} & \multicolumn{2}{c}{Linux, RTX A6000, Paris} & \multicolumn{2}{c}{Mac M2, MPS, local} & \multicolumn{2}{c}{Mac M2, MPS, SD, remote} \\
\hline
Model & Diff. & CD & Diff. & CD & Diff. & CD & Diff. & CD \\

\hline
MAX/MSP $\to$ server        &  17       &  17       & 188       & 188       &  20         &  20         & 107         & 107 \\
CAE encode                  &  40       &  40       &  52       &  52       &  55         &  55         &  55         &  56 \\
Sampling ($N$ fwd.\ passes) & 1175\,(10) & 130\,(2) & 480\,(10) & 88\,(2)   & 1072\,(10)  & 146\,(2)    & 1072\,(10) & 146\,(2) \\
CAE decode                  & 141       & 150       &  72       &  72       &  89         &  90         &  89         &  90 \\
Server $\to$ MAX/MSP        &  25       &  25       & 189       & 189       &  20         &  20         & 107         &  107 \\
\hline
Full cycle                  & 1398       & 362       & 981       & 589       & 1256       & 331         & 1434       & 506 \\
RT                          & $\checkmark$ & $\checkmark$ & $\checkmark$ & $\checkmark$ & $\checkmark$ & $\checkmark$ & $\checkmark$ & $\checkmark$ \\
\hline
\end{tabular}}
\end{table*}

\subsection{RTAP Setup}
\label{sec:system_config}

In accordance with our generative models, the default settings of the RTAP system are $T{=}6$\,s, $r{=}0.25$, giving a step size of $T{\cdot}r{=}1.5$\,s (with 44.1 kHz giving 66,150 samples), and $w{=}1$ (Look-ahead mode). OSC audio data is transmitted in chunks of 4,410 samples (0.1\,s), yielding 15 packets per step — a chunk size chosen to balance throughput with reliability. A \texttt{Fade} of 0.02\,s ($\approx$882 samples, $\approx$20\,ms) is applied at the write boundary to suppress clicks. The Max/MSP host buffer size is set to 64 samples ($\approx$1.45\,ms) to minimise client-side audio latency, since no heavy computation occurs on the Max side.

The system was tested across four deployment configurations: (1) a local Windows machine (Windows 10, NVIDIA GeForce RTX 2070 Max-Q) acting as both client and inference server; (2) a Windows client in San Diego, USA connected remotely to a Linux server (kernel 5.10, NVIDIA RTX A6000, 48\,GB) at IRCAM in Paris, France over a standard internet connection; (3) a local Apple M2 Mac (macOS 14, MPS backend) running the inference server natively; and (4) a Windows client connected remotely to an Apple M2 Mac server (macOS 14, MPS backend) over SSH. These cover a range of practical scenarios from high-performance remote inference to local laptop deployment. In practice, configuration~(2) — a powerful remote GPU server accessed over the internet — has been our most frequently used deployment.

\section{Results}
\label{sec:results}

\subsection{Generative Model Performance}

Fig.~\ref{fig:sweep_analysis} summarises performance of our diffusion model and consistency distillation (CD) model across COCOLA, Beat $F_1$, and FAD, compared against the StreamMusicGen's online decoder and offline baselines (Prefix Decoder, StemGen), as a function of the net look-ahead $T \cdot r \cdot w$ — the effective time distance between the current playback position and the start of the predicted window. The comparison is indicative rather than direct due to differences in instrument choices, audio codec, sample rate, and streaming paradigm. However, the results still provide insights into the relative performance of our models across different generative scenarios.

\textbf{Musical coherence (COCOLA).}
Both our models follow the same trend as the StreamMusicGen Online Decoder: coherence increases as more context is available, peaking in the Retrospective zone and decreasing in the Immediate and Look-ahead zones. In the Retrospective zone ($w{=}{-1}$), our diffusion model performs comparably to the Online Decoder and approaches the offline Prefix Decoder at the largest step size $r{=}1$. It also performs on par with the ground-truth (GT) ceiling, indicating high generation quality. The difference in GT COCOLA between our models and the baselines reflects the difference in instrument scope: the score is generally higher across all Slakh instruments than for our four-stem subset, likely due to the greater harmonic diversity present in the full-instrument setting. In the Immediate zone ($w{=}0$), our models show lower coherence. Finally, in the Look-ahead zone ($w{=}1$), our models exceed the StreamMusicGen Online Decoder at $r{=}0.125$ (0.75\,s look-ahead) and $r{=}0.25$ (1.5\,s look-ahead). (Note that these two are the only feasible ratios in our configuration, as $r \geq 0.5$ would place the prediction window entirely outside the context when $w{=}1$.) Achieving higher COCOLA in the Look-ahead zone despite operating on a lower-scoring instrument subset underscores our model's advantage over the baseline in this regime. The CD model follows a similar trend but with slightly lower overall coherence, suggesting that distillation may reduce sensitivity to fine-grained musical relationships. Nevertheless, the CD model still marginally outperforms the baseline model at the 1.5\,s look-ahead window.

\textbf{Rhythmic alignment (Beat $F_1$).}
Beat alignment shows a similarly strong dependence on context availability. In the Retrospective zone, our diffusion model largely surpasses the Online Decoder and approaches the ground-truth (GT) ceiling, even matching it at $r{=}1$. Unlike COCOLA, the GT Beat $F_1$ ceiling is higher for our four-stem subset than for the full Slakh instrument set, as bass, drums, guitar, and piano are inherently rhythmically active — inflating beat alignment scores regardless of generation quality. Notably, our model approaching the GT ceiling in the Retrospective zone is a strong result, while the baselines do not achieve the same relative to their own GT ceiling. In the Immediate zone, there is a sharp drop in score, yet our diffusion model still outperforms the Online Decoder. In the Look-ahead zone, scores drop further — though less steeply — and our model continues to surpass the baseline; however, given the rhythmically biased instrument set, the remaining margin above the random-pairing floor is modest, suggesting that rhythmic coherence in the Look-ahead regime remains a challenge. The CD model consistently lies below the diffusion model across all zones, still outperforming the baseline but approaching the random-pairing floor in the Look-ahead zone.

\begin{table*}[t]
\centering
\caption{Full-cycle inference timings (ms) for both models across step ratios $r$, measured on a dedicated NVIDIA RTX A6000 GPU with no competing processes. MAX$\to$server: OSC transfer of a $T{\cdot}r$ audio chunk from MAX/MSP to the Python server; enc.$+$sample: CAE encoding plus iterative sampling (10 forward passes for Diffusion, 2 for CD); dec.: CAE decoding to waveform; server$\to$MAX: OSC return of the generated $T{\cdot}r$ audio chunk back to MAX/MSP. RT indicates whether the real-time constraint $d < T{\cdot}r$ is satisfied ($\checkmark$) or not ($\times$). $^\dagger$Theoretical minimum step ($r_{\min}{=}1/64$, one latent time step).}
\vspace{1mm}
\label{tab:latency_r}
\resizebox{0.7\textwidth}{!}{%
\begin{tabular}{clccccccc}
\hline
Model & $r$ & $T{\cdot}r$ & MAX$\to$server & enc.$+$sample & dec. & server$\to$MAX & Total & RT \\
\hline
Diff. & 0.250         & 1500 & 188 & 532 & 72 & 189 &  981 & $\checkmark$ \\
      & 0.125         &  750 &  145 & 532 & 67 &  145 &  889 & $\times$ \\
      & 0.0625        &  375 &  145 & 532 & 64 &  145 &  886 & $\times$ \\
      & $1/64^\dagger$&   94 &  145 & 532 & 64 &  145 &  886 & $\times$ \\
\hline
CD    & 0.250         & 1500 & 188 & 140 & 72 & 189 &  589 & $\checkmark$ \\
      & 0.125         &  750 &  145 & 140 & 67 &  145 &  497 & $\checkmark$ \\
      & 0.0625        &  375 &  145 & 140 & 64 &  145 &  494 & $\times$ \\
      & $1/64^\dagger$&   94 &  145 & 140 & 64 &  145 &  494 & $\times$ \\
\hline
\end{tabular}
}
\end{table*}

\textbf{Audio quality (FAD).}
FAD follows the same trend. All models degrade in quality from the Retrospective to the Immediate and Look-ahead zones. However, this degradation is more pronounced in our models than in the baseline: they outperform the baseline in the Retrospective zone but underperform in the remaining zones, with the CD model further disadvantaged by the reduced number of sampling steps.

Overall, our models demonstrate strong accompaniment generation quality, approaching the ground-truth ceiling and achieving low FAD scores in the Retrospective zone. Both our models and StreamMusicGen similarly struggle in the Look-ahead zone, suggesting that real-time accompaniment generation with look-ahead remains an open challenge across paradigms.

\subsection{Real-Time System Performance}
\label{sec:latency}

We measured the end-to-end processing time of both the diffusion and CD models as deployed with the MAX/MSP client–server system. Each complete inference cycle consists of five sequential stages: \textbf{(1)} \textbf{\emph{MAX/MSP$\,\to\,$server}}, transfer of a new $T \cdot sr \cdot r$ audio chunk from the MAX/MSP client to the Python inference server via OSC, where it is appended to the server's rolling context tensor of length $T \cdot sr$; \textbf{(2)} \textbf{\emph{CAE encoding}}, the full context tensor is projected into the $T_z {\times} F_z$ latent space by the frozen Music2Latent encoder; \textbf{(3)} \textbf{\emph{sampling}}, iterative denoising for the diffusion or CD with respective number of steps; \textbf{(4)} \textbf{\emph{CAE decoding}}, reconstruction of the generated waveform from the last $T_z \cdot r$ predicted latent frames, yielding $T \cdot sr \cdot r$ audio samples; and finally, \textbf{(5)} \textbf{\emph{server$\,\to\,$MAX/MSP}}, return of the newly generated $T \cdot sr \cdot r$ audio samples to the MAX/MSP client via OSC.

Table~\ref{tab:latency} reports mean stage timings for both models at $r{=}0.25$ ($T \cdot sr \cdot r{=}66{,}150$ samples, $T{\cdot}sr{=}264{,}600$ samples at 44.1\,kHz, and in the latent space $T_z{\cdot}r{=}16$ latent frames) across all the configurations described in Sec.~\ref{sec:system_config}. The dominant difference between timing of the modes is in the sampling stage: the diffusion model runs 5 denoising steps with 2 resamples per step (10 forward passes total), while the CD model uses 2 consistency steps — yielding approximately a $5{\times}$ reduction in sampling time consistently across all platforms. As expected, transfer times depend heavily on network topology: local configurations show send and receive times of $\sim$20\,ms, while remote connections are significantly longer — the San Diego–Paris link (config 2) incurs nearly twice the transfer latency of a same-city remote connection (config 4). Compute stages (CAE encoding and sampling) are platform- and hardware-dependent, with the Linux RTX A6000 (config 2) significantly faster than local and M2 MPS backends (configs 1, 3, 4). At $r{=}0.25$, all configurations satisfy the real-time constraint ($d < T{\cdot}r = 1.5$\,s). We note that the reported timings are mean values and can fluctuate depending on GPU load and network conditions.

To further investigate the effect of step size on end-to-end cycle time, we measured both models across four step sizes ($r \in \{1/64,\, 0.0625,\, 0.125,\, 0.25\}$) using configuration~(2). $r{=}1/64$ represents the absolute lower bound set by the latent temporal resolution (since Music2Latent compresses $T{=}6$\,s to $T_z{=}64$ frames, $r=1/64$ represents the minimal step scenario where the window advances by a single latent frame per cycle). As shown in Table~\ref{tab:latency_r}, the CAE encoder and sampling stages are constant regardless of $r$ — a fixed compute cost that can only be reduced by changing the model. In contrast, the transfer and CAE decoder times are $r$-dependent, as we decode only the newly generated $T \cdot r$ portion of the latent output. Both decrease with $r$ but plateau below $r{=}0.125$, where transfer times plateau due to the network round-trip floor (San Diego–Paris), and the CAE decoder similarly saturates as the latent chunk size shrinks. As a result, the diffusion model fails to satisfy the real-time constraint at $r{=}0.125$ (889\,ms vs.\ 750\,ms threshold), while the CD model still satisfies it (497\,ms), but neither model satisfies the constraint at smaller values of $r$.

Unlike the remote setting above, in a local deployment — where the inference server runs on the same machine as the MAX/MSP client (configurations~(1) and~(3)) — transfer times scale proportionally to chunk size without plateauing. The total cycle time can then be decomposed as $d(r) = d_{\text{compute}} + c \cdot r$, where $d_{\text{compute}}$ collects all $r$-independent costs (CAE encoding, sampling, and decoding), and $c \cdot r$ captures the transfer overhead scaling linearly with chunk size. The real-time constraint $d(r) < T \cdot r$ then gives the minimum feasible step ratio:
\begin{equation}
    r^* = \frac{d_{\text{compute}}}{T - c},
\end{equation}
where $c$ is the transfer coefficient (ms per unit $r$), estimated from local measurements. In our experiments with configurations~(1) and~(3), testing at $r{=}0.25$ and $r{=}0.125$, local transfer takes ${\approx}20$\,ms and ${\approx}10$\,ms respectively, giving $c \approx 80$\,ms per unit $r$. To estimate the theoretical best case, we use the RTX A6000 compute costs from configuration~(2) ($d_{\text{compute}} \approx 596$\,ms for diffusion, $\approx 204$\,ms for CD) — the fastest GPU available — combined with the local transfer coefficient $c$ and $T{=}6$\,s, yielding $r^*\approx 0.100$ for the diffusion model — placing the minimum feasible latent-aligned step at $r{=}7/64\approx 0.109$ ($T{\cdot}r\approx 0.656$\,s) — and $r^*\approx 0.034$ for the CD model, placing it at $r{=}3/64\approx 0.047$ ($T{\cdot}r\approx 0.281$\,s). These estimates indicate how fine a step size could theoretically be supported with the current models under ideal local conditions. However, these step sizes were not used during training, so generation quality at such granularities is not guaranteed; retraining with finer masking ratios and empirical validation are left for future work.

\section{Conclusion}
\label{sec:conclusion}

We present a framework for real-time human--AI musical co-performance combining a latent diffusion model with a sliding-window look-ahead inference paradigm, accelerated via consistency distillation, and deployed through a low-latency client--server system interfaced via RTAP, a musician-facing MAX/MSP patch. In this work, we establish that the central challenge of real-time accompaniment generation under non-negligible inference latency is that the system must anticipate future audio segments by pre-generating them --- decoupling generation from playback and necessitating models both trained and inferred under a look-ahead paradigm that explicitly supports partial musical context. A central contribution of this work is the proposed sliding-window look-ahead paradigm with dedicated masked context conditioning, which directly addresses this constraint and constitutes a principled inference framework transferable to any model employing a similar inpainting/outpainting conditioning scheme. Since model speed determines the feasible look-ahead window, we apply consistency distillation to our base diffusion model, which proves effective at reducing inference latency without substantial loss in generation quality, enabling real-time operation at step sizes unachievable by the base model. Evaluated against StreamMusicGen, both models exhibit the same qualitative trend --- quality peaks in the Retrospective regime and degrades with increasing Look-ahead --- while achieving marginally better scores across all regimes, confirming that high-quality look-ahead audio generation largely remains an open challenge. Another central contribution is the RTAP system: a model-agnostic, low-latency client--server interface validated across local and remote deployments on multiple platforms, designed to accommodate future models adapted to the streaming protocol. These results offer both a concrete operational system for practitioners and a principled foundation for future research, motivating further work on model acceleration, architectural efficiency, and fine-grained step-size training to push the boundaries of real-time co-performance.

\section{Acknowledgments}

We thank the Institute for Research and Coordination in Acoustics and Music (IRCAM) and Project REACH: Raising Co-creativity in Cyber-Human Musicianship for their support. This project received support and resources in the form of computational power from the European Research Council (ERC REACH) under the European Union’s Horizon 2020 research and innovation programme (Grant Agreement 883313).

\bibliographystyle{IEEEbib}
\bibliography{refs}

\end{document}